# Serving economic prosperity: economic impact assessments (EIA) on Earth observation-based services and tools by SERVIR

Reetwika Basu[1,2], Eric Anderson[1,3], Chinmay Deval[1,2], Kelsey Herndon[1,2], Amanda Markert[1,2], Lena Pransky[1,2], Emil Cherrington[1,2], Aparna Phalke[1,2], Alqamah Sayeed[1,4]

1 SERVIR Science Coordination Office, NASA Marshall Space Flight Center, Huntsville, AL, United States
2 Earth System Science Center, University of Alabama in Huntsville, Huntsville, AL, United States
3 Earth Science Branch, NASA Marshall Space Flight Center, Huntsville, AL, United States
4 Universities Space Research Association, Huntsville, AL, United States

## Abstract

In an era where informed decision-making is paramount for sustainable development and effective resource management, the role of Earth observations (EO) in shaping economic landscapes cannot be overstated. EO, facilitated by satellites, sensors, and data analytics, is a cornerstone for evidence-based policymaking, risk mitigation, and resource allocation. From monitoring environmental changes to forecasting disasters, these observations equip governments, businesses, and communities with information to anticipate challenges, capitalize on opportunities, and foster resilience in the face of uncertainty. SERVIR, a joint program of the U.S. Agency for International Development (USAID) and NASA, partners with experts across the globe to use EO to address environmental and climate issues. This paper presents a comprehensive survey of relevant economic impact assessment (EIA) work, summarizes SERVIR's potential interests in EIA, and identifies how and where EIA could improve how SERVIR quantifies and communicates the impact of its services.

This paper is aimed at answering several key questions. First, what are the benefits of using EO in decision-making? Second, which SERVIR services and tools are most suitable for an EIA, and what are the gaps and opportunities that exist in SERVIR services? Third, what are the typical methods used for EIA, and what potential methods could be employed by SERVIR?

To understand the benefits of EO, we compiled the findings of case studies conducted by other groundbreaking EIA efforts, such as the Valuation of Applications Benefits Linked with Earth Science (VALUABLES) consortium. To understand the relevance of EIA for SERVIR's services, we held discussions network-wide and subsequently assessed the feasibility of conducting EIAs based on the available data and methodologies. Based on the findings, we propose conducting initial SERVIR EIAs related to 1) the economic impacts of the Smoke Watch tool on health and greenhouse gas emissions in Thailand, and 2) the impact of subseasonal-to-seasonal (S2S) forecasting on agriculture outcomes in Ghana. Additionally, we introduce an ongoing collaboration with the National Oceanic and Atmospheric Administration (NOAA) to conduct an EIA of the Group on Earth Observations Global Water Sustainability (GEOGLOWS) streamflow prediction services in Ecuador, with potential replication in other countries.

This work provides a comprehensive examination of case studies from SERVIR and the VALUABLES consortium. The questions identified and synthesis help pave the way for future potential collaborations with the Collaborative Network for Valuing Earth Information (CONVEI). The methods and key findings are summarized, from which recommendations on improving impact assessments are proposed. Future work aims to enhance the value of EO services offered by SERVIR and other similar programs, and strengthen the emerging field of EIA for the valuation of EO-driven services.

# 1. Introduction

Economic Impact Assessment (EIA) refers to a broad set of methods used to identify and analyze the impacts of existing projects, programs, or policies on the economy of the application region. It can also identify potential impacts of a new policy or program yet to be implemented. It is a decision-making tool that can be used to identify the applicability and feasibility of a project and show project impacts on factors like production, jobs, income, livelihood, and costs. It can also analyze ecological changes caused by policies and programs.

The use of Earth observation (EO) data can have significant economic impacts across various environmental and climate issues (Anderson et al., 2017; Kansakar & Hossain, 2016). EO data can be used to monitor crop health, crop yield, consumptive water use, and evapotranspiration. It can be used to detect conditions conducive to early-onset crop disease and pest infestation, to monitor and forecast adverse weather conditions, potential flood impacts, and to help farmers prepare for and adapt to changing water availability and climate conditions. EO data can also be used to monitor deforestation, illegal mining, sustainable forestry management, water quality, and the impact of climate change on water availability.

Several methods can be used to conduct an EIA of EO applications. Cost-benefit analysis (CBA), return on investment (ROI), environmental impact assessment, randomized controlled trials (RCTs), econometric models, and general and partial equilibrium (GE and PE) models are the most widely used economic assessment methods. Readers are encouraged to refer to the Appendix for definitions of all the economic assessment methods discussed in this white paper. Impact evaluation, in general, is at the core of the United States Agency for International Development's (USAID) work, primarily through monitoring, evaluation, and learning (MEL) frameworks. The National Aeronautics and Space Administration (NASA) and other agencies such as the US Geological Survey (USGS) and the National Oceanic and Atmospheric Administration (NOAA) are also adding to the growing body of knowledge on investigating the socio-economic impacts of satellite data and derived products.

SERVIR is a joint initiative of NASA and USAID. It leverages satellite data and geospatial technologies to enhance weather and climate resilience, agriculture and food security, water security, ecosystem and carbon management, and air quality and health. SERVIR activities are implemented by a global network of applied research organizations across four regional hubs in Africa, Asia, and Latin America. These organizations collaborate with communities of all sizes, from sub-national to national to regional, to develop locally led solutions (referred to as SERVIR "services") tailored to address the unique challenges faced in each region.

EIA is imperative to understanding the socioeconomic impacts of SERVIR services. EIA goes beyond monetary assessments to encompass various impacts such as changes in crop yield, acres of forests preserved, lives saved from reduced air pollution, reduced carbon emissions, and time and resources saved with improved information. SERVIR aims to use EIA to understand the impact of its services across all hub regions. The objective of this paper is to identify questions about SERVIR services that could be addressed through EIA, and then narrow down one or two services for in-depth assessment. Subsequently, service area-specific roadmaps will provide strategic tools and guidelines for future EIAs. SERVIR aims to build on this initial assessment to eventually integrate EIA into all hub regions and thematic areas, enhancing service design and delivery. SERVIR also seeks to collaborate with other stakeholders in the emerging field of EIA for EO-driven services valuation and collectively advance interdisciplinary understanding of impacts.

## 1.1. *History of SERVIR's impact assessment and evaluation*

While economic methods have not traditionally been central to SERVIR's services, its Service Planning Toolkit and MEL approaches have paved the way for EIA within the SERVIR network (USAID & NASA, 2021). Figure 1 provides a timeline of evaluations and impact assessments conducted within SERVIR, along with several other milestones in NASA's impact assessment work related to EO resulting in Impact Assessment from SERVIR Science Coordination Office (SCO).

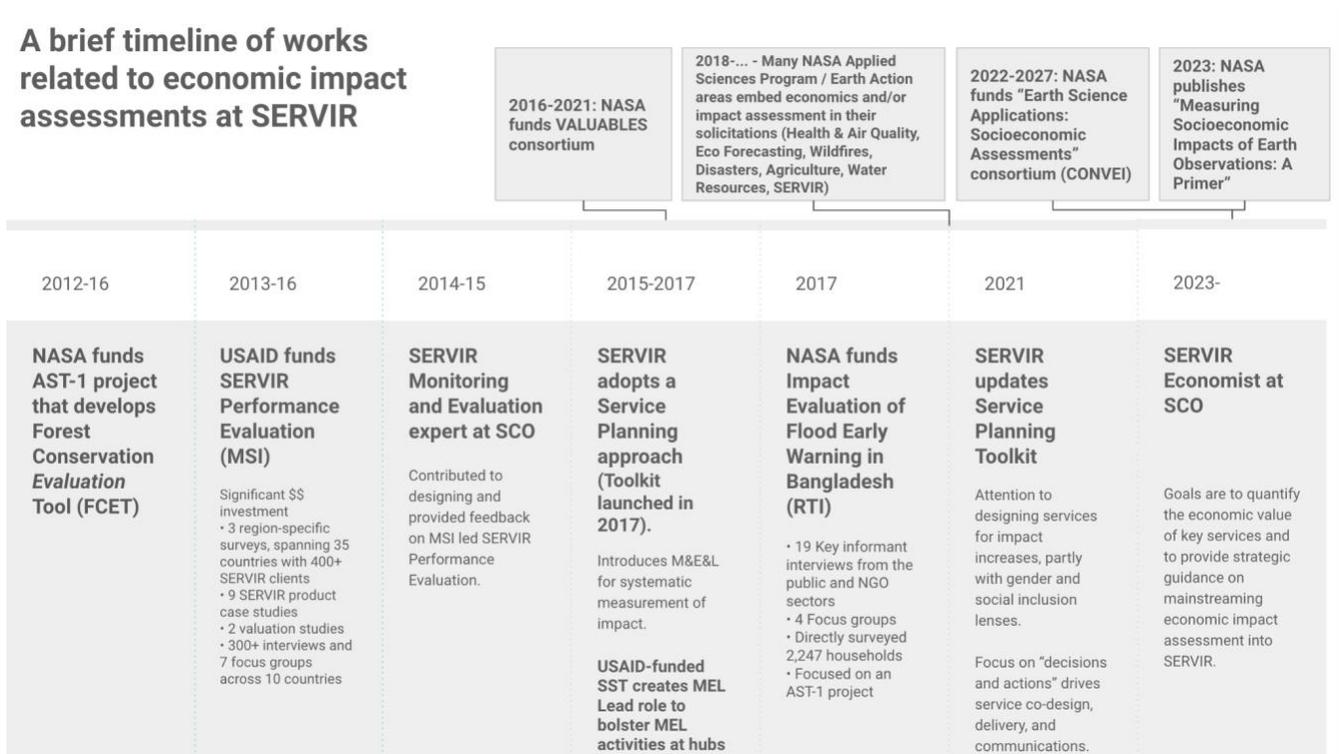

**Figure 1:** A brief history of evaluations and impact assessments at SERVIR. An AST (Applied Science Team)-1 project developed a Forest Conservation Targeting Tool (Blackman et al 2019; https://fctt.servirglobal.net/) and a Forest Conservation Evaluation Tool (https://fcet.servirglobal.net/). SERVIR conducted a program-wide performance evaluation (Morrison et al 2018a; Morrison et al 2018b; USAID 2018). Another targeted study focused on floods in Bangladesh (RTI International, 2018; as summarized by French & Hernandez 2018). NASA's primer on measuring socioeconomic impacts of Earth observations lay a framework for evaluating and communicating benefits of EO investments (NASA 2023).

This paper is the first step to summarize the SERVIR network's interest and readiness to use EIA approaches. The following sections summarize previous impact evaluations conducted by VALUABLES and SERVIR, summarize findings from discussions network-wide, and assess the readiness of services and thematic areas to carry out EIA.

## 2. Case Studies from VALUABLES and SERVIR

EO provides detailed data on Earth's systems, aiding in tracking natural impacts and informing prioritized actions. Examples include evaluating conservation programs' impacts on biodiversity, identifying methane emissions sources, and forecasting wildfires for emergency planning. EO not only democratizes knowledge but also raises questions on its optimal utilization for a sustainable and equitable future. Considering impact evaluations for using EO, the SERVIR performance evaluation was pivotal in shaping the program's strategy and service planning approach today, but it focused little on *economic* impact assessment. Thus we have much to learn from the following case studies of impact analyses conducted by our parent agencies, NASA and USAID. This includes impact assessments by the NASA-supported VALUABLES consortium.

*Additional background*

VALUABLES is a collaboration between Resources for the Future (RFF) and NASA to assess the benefits of using satellite data in decision-making (https://www.rff.org/valuables/, supported through NASA Cooperative Agreement NNX17AD26A with RFF). The consortium involved economists, decision scientists, NASA experts, and Earth science professionals seeking to quantify the value of satellite information for societal benefit using the science of the Value of Information (VOI). The case studies focused on evaluating the value of satellite data for activities such as protecting endangered species, informing post-wildfire responses, enforcing air quality standards, and improving flood and drought forecasts.

CONVEI (Collaborative Network for Valuing Earth Information) is a new collaboration between the World Wildlife Fund (WWF), NASA, NOAA, and USGS focused on studying how information, mainly EO, can enhance decision-making for sustainable development (https://www.worldwildlife.org/projects/convei-collaborative-network-for-valuing-earth-information, funded through NASA SMD Grant 80NSSC23K0914 with WWF). While VALUABLES primarily used the VOI approach to analyze impacts, CONVEI will be using new methods for impact evaluation, drawing from fields including economics, machine learning, behavioral psychology, and cultural anthropology. These methods will explore the societal valuation of information services related to agriculture, water resources, natural resources and disaster management, climate resilience, air quality and health, and environmental and social justice.

From 2013-2016, USAID's Bureau for Economic Growth, Education, and Environment (E3) commissioned an evaluation of SERVIR to understand its effect across the world. The study aimed to understand how SERVIR services affected its users and beneficiaries. The effects were segregated into environmental, economic, and social well-being. SERVIR impact evaluations incorporated a limited amount of economic methods, and the maturity of SERVIR products, many being at nascent stage, at the time made it difficult to assess the impact accurately. Nevertheless, the results showed that the products did demonstrate substantial social and economic value, and the few economic models that were used revealed the aspects of the products that were most valuable to users. Based on the recommendations from the Evaluations Report, it is essential to understand the extent and impact of using SERVIR products that are already deployed to allow them to be used to their maximum potential (Morrison et al 2018a; Morrison et al 2018b; USAID 2018).

### 2.1. *Summary of VALUABLES Impact Assessments & SERVIR Reports*

This section summarizes 15 impact assessments performed by VALUABLES, SERVIR's impact evaluation of flood early warning systems in Bangladesh, and the SERVIR performance evaluation (PE). Some studies focus on potential economic impacts should EO be incorporated into existing decision-making processes, while others estimate the actual economic impacts of EO-related interventions. This section offers some lessons learned from each of the studies and an understanding of the methods that can be used for impact evaluations.

**Table 1:** The objectives, approaches, and key findings from the impact assessments related to EO

| OBJECTIVE | APPROACH[1] | RELEVANT KEY FINDINGS |
|---|---|---|
| **VALUABLES Impact Assessments** | | |
| What is the expected reduction in whale deaths per year to meet the established annual Potential Biological Removal (PBR) at least costs? (Geographic Location: Eastern North Pacific Ocean)(Bernknopf et al., 2019) | ● Decision model<br>● VOI | *Potential impacts identified.*<br>● In situ data on the number of strikes per year varies from remotely sensed data.<br>● Use of remotely sensed data could save $21 million for 7.8 fatal strikes, rising to $332 million for 5.3 fatal whale strikes. |
| Identify how much of the U.S. population lies outside the area where air quality is appropriately monitored. Are vulnerable populations misclassified [and can satellite-based information help improve the classification]? (Geographic Location:Continental U.S.)(Sullivan & Krupnick, 2018) | ● DID model<br>● VSL | *Actual impacts identified.*<br>● 24.4 million people lived in misclassified areas, and out of them 10.9 million live in counties with no air quality monitors.<br>● Illinois, California, and Texas were observed to have the largest population of unmonitored and misclassified people.<br>● 23.2 million people living in non-attainment areas are predicted to live in urban environments more than residents in properly classified attainment areas (97 percent vs. 78 percent).<br>● Using EPA standard VSL of $9 million, the social cost of the excess mortality was observed to be $24.5 billion per year.<br>● Total excess mortality of misclassification would be 5452 with a social cost of $49 billion. |
| How does a mobile phone-based information dissemination campaign on infectious disease risks impact household behavior? (Geographic Location:Rural Bangladesh)(Pakhtigian et al., 2022) | ● DID model<br>● Linear regression model | *Actual impacts identified.*<br>● User-directed information was found to be effective in promoting safe water hygiene and sanitation behaviors. For instance, pond water was no longer used for drinking purposes.<br>● CholeraMapp users washed their hands more frequently than the other app users and relied on the app not only for early warning systems but also as reference source for households experiencing Cholera or Cholera like illness.<br>● CholeraMap users reported a 9.3 percent and 7.7 percent increase in their confidence to respond to environmental risk. |

---

[1] Refer to Key Concepts for detailed description

| OBJECTIVE | APPROACH[1] | RELEVANT KEY FINDINGS |
|---|---|---|
| How to use the social cost of carbon to compute the Value of Information provided by projected climate observation system improvements? (Cooke et al., 2013) | • NVP<br>• Integrated assessment models<br>• VOI | *Potential impacts identified.*<br>• The value of information of an advanced climate observing system using the CLARREO example was large relative to cost.<br>• The current climate observations cost in the US is approximately $2.5 billion per year with international efforts of around $5 billion per year. |
| Develop and demonstrate a framework to estimate the value of satellite data for identifying the presence of Harmful Algal Blooms (HABs) in surface waters (Geographic Location:100 lakes in California) (Newbold et al., 2022). | • RUM<br>• Revealed preference econometric model | *Potential and Actual impacts identified.*<br>• Recreator's preferences regarding visiting a recreation park with HABs was found to be heterogeneous.<br>• The total number of visits to the 100 lakes in California selected for the study, was 17.3 million.<br>• The value per trip was estimated to be roughly between $8 to $10.<br>• The estimated willingness-to-pay was roughly $15 for the top 50 most severely impacted lakes.<br>• The total value of a perfect early warning system of HABs for the 100 chosen lakes and the used dataset within the time frame mentioned above was estimated to be $2.46 million. |
| What is the value of a crop that could be harvested by reduced uncertainty in US corn and soybean crop yields? (Cooke et al., 2019) | • VOI | *Potential impacts identified.*<br>• 30 percent reduction in the weather component leads to US Consumer Surplus of $1.44 billion.<br>• About half of the value of perfect weather information is obtained by reducing the weather uncertainty by 30 per cent. |
| What is the socio-economic benefit of using satellite studies in halting transmission of Polio in Nigeria? (Mariel et al., 2023) | • VOI<br>• Survey<br>• Difference in magnitude of benefit was compared to a counterfactual scenario where satellite products were not used | *Actual impacts identified.*<br>• Using satellite data, in 2014, the net socio-economic benefit was between $46 - 153.9 million.<br>• The population was more vulnerable than what was captured in census data.<br>• 56 to 172 cases of polio were avoided using satellite data.<br>• 10 other countries in Africa used the same process to support vaccination and health efforts after observing the results. |
| What are the cost-savings realized from using Burn Area Emergency Response (BAER) to identify post wildfire threats to human life, safety, property, and critical natural and cultural responses? (Case Study: 2003 El Complex Fire at the Boise National Forest of Idaho) (Bernknopf et al., 2019) | • VOI | *Actual and Potential impacts identified.*<br>• The team prioritized treatment areas based on burn severity, reducing the treatable acreage from 16,000 to 2,000-4,000 acres using Landsat-derived information.<br>• Both Landsat and commercial satellite data inputs are more cost-effective than relying on surveys in helicopters.<br>• There were substantial cost savings over a period of five years. |

| OBJECTIVE | APPROACH[1] | RELEVANT KEY FINDINGS |
|---|---|---|
| What is the value of information from NASA's Gravity Recovery and Climate Experiment (GRACE) satellite on groundwater storage when the information is fed into the U.S. Drought Monitor (USDM)? (Bernkopf et al., 2018) | ● Multi-model framework with a Bayesian updating process.<br>● Stakeholder listening sessions.<br>● Econometric analysis | *Potential impacts identified.*<br>● When compared to current drought monitor and relief programs, relying solely on GRACE-DA indicators would have significantly altered funding allocations, benefiting certain states, and disadvantaged others?<br>● While acknowledging the hypothetical nature of the study, the authors suggest that GRACE-DA has the potential to reduce uncertainty in drought understanding, influencing policy decisions for social benefits |
| How conflicting scientific views are sought out or generated from unwanted scientific advice? [An enhanced Earth Observation System (EOS) component called CLARREO is compared againstInfrared Atmospheric Sounding Interferometer(IASI), Atmospheric Infrared Sounder (AIRS), Cross-track Infrared Sounder (CrIS), Clouds and the Earth's Radiant Energy System (CERES) and Cloud Radiative Forcing (CRF) for global surface temperature rise and decadal percentage change.] (Cooke & Wielicki, 2018) | ● Bayesian net simulations | *Characterization of uncertainty; potential impacts identified.*<br>● Using only enhanced systems for measuring temperature significantly reduces the uncertainty of Equilibrium Climate Sensitivity (ECS) as compared to old systems.<br>● Average posterior standard deviation of ECS is 0.49 with the enhanced system. |
| Estimation of counterfactual deforestation and carbon emission by assessing what would have occurred without the (Real Time Deforestation Detection System) DETER satellite system. [Focused on the legal Amazon region of Brazil from 2000 to 2015]. (Mullan et al., 2022) | ● Estimation models<br>● Causal mediation analysis | *Actual impacts identified.*<br>● Annual deforestation is significantly higher under counterfactual scenarios without DETER.<br>● Avoided deforestation is approximately 467 to 471 km2. This indicated avoiding almost 12 billion tons of carbon emissions. |
| Assessing the recent advances in modeling ice sheets and the need to have more accurate predictions (Geographic Locations: Greenland, West Antarctic, East Antarctic) (Bamber et al., 2019). | ● Structured expert judgment (SEJ) exercise<br>● Classical model decision maker | *Characterization of uncertainty*<br>● By 2050, the median ice sheet contribution to sea level rise was expected at 10-12 cms. |
| Monetize the value of measurements of equilibrium climate sensitivity using social cost of carbon (Cooke et al., 2019) | ● DICE Model<br>● Real Option Value (ROV) | *Potential impacts identified.*<br>● Expected net benefits for different Equilibrium Climate Sensitivity (ECS) values, discount rates and trigger values were calculated.<br>● There should be more investment toward cloud radiative effect (CRE) relative to global surface temperature (GSF) |
| What is the importance of using mid-level resolution | ● VOI | *Actual impacts identified.*<br>● The Burn Area Reflectance Classification (BARC) |

| OBJECTIVE | APPROACH[1] | RELEVANT KEY FINDINGS |
|---|---|---|
| satellite imagery (Landsat), for efficient and cost-effective post wildfire response activities? (Bernknopf et al., 2020) | | map derived from Landsat aids in assessing burn severity, and its accuracy is enhanced through in-situ sampling and adjustments.<br>● Savings range from US$11,157 to US$51,418 per incident for the reference case.<br>● Extrapolating over five years, the estimated savings for using Landsat imagery in BARC production and BAER response range from US$7.5 million to US$35 million. |
| **SERVIR's Impact Assessments** | | |
| Evaluation of flood early warning system in Bangladesh based on JASON-2 satellite providing 3 to 5 days of advanced flood warnings (RTI International, 2018). | ● Institutional and field surveys<br>● Economic valuation models | *Actual and Potential impacts identified.*<br>● Reduced flood losses in 3 days warning by 5.5% saving $73.6 million. An 8-day warning increased savings by $11.6 million |
| *The following case studies are from Reports 1 (Morrison et al 2018a) and 3 (Morrison et al 2018b), with economic impacts listed below:* | | |
| Are SERVIR's products and applications being used in decision-making contexts? How (1) for hydrology (floods, etc.); (2) for land use, biodiversity and ecosystems; and (3) for disasters (fires, droughts, frost, etc.)?<br><br>What are the measurable impacts, both intended and unintended, of SERVIR's products and applications on the relevant societal benefit areas themselves in the countries where we work?<br><br>What is the value calculated as benefits of SERVIR's capacity building, science applications, data sharing efforts and global network? | ● Cross-case valuation of nine case studies across SERVIR regions<br>● Economic valuation models | *Actual and Potential impacts identified.*<br>● SERVIR products have demonstrated social and economic value for thousands of households and administrative value for government service providers.<br>● Greater clarity on which aspects of SERVIR products were most economically valuable to users.<br>● The products have added value in difficult-to-access terrains. |
| *Frost Mapping, Monitoring, and Forecasting System in Kenya* | ● Economic valuation model - Measurement of loss avoided | *Potential impacts identified.*<br>● The SERVIR frost monitoring tool which provides 3 day warning could save a Kenyan tea farmer $80.47, the equivalent of 25 days of household food spending, almost a full year of a child's school tuition, or a full year of household health spending |
| *Forest Fire Monitoring in Guatemala* | ● Economic valuation model - contingent valuation methods | *Actual impacts identified.*<br>● The users of SERVIR forest fire hotspot monitoring tool in Guatemala were willing to pay $78 per year on average for daily access because of the tool's frequency and reliability of reporting. |

| OBJECTIVE | APPROACH[1] | RELEVANT KEY FINDINGS |
|---|---|---|
| *Coupled Routing and Excess Storage Tool (CREST) Hydrological Suite* | ● Economic valuation model (limited information available) | *Actual impacts identified.*<br>● CREST hydrological modeling tool portrayed informed economic viability assessment for infrastructure projects. There were indications of loss and property damages avoided. |
| *Ocean Algal Bloom Monitoring for Mesoamerica - designated location: El Salvador* | ● Economic valuation model (limited information available) | *Actual impacts identified.*<br>● Long term increased effectiveness of algal bloom monitoring services have been observed to have strengthened market confidence. |
| *Water Quality Monitoring for Lake Atitlán, Guatemala* | ● Surveys, focused group interviews | *Actual Impacts identified.*<br>● The awareness brought to the government regarding water quality in the lake significantly influenced decision-making regarding local, national and international level fundings. |
| *Rapid Response Mapping for Natural Disasters in the Himalayas* | ● Surveys, focused group interviews | *Actual Impacts Identified.*<br>● The EO data for Seti and Sunkoshi Decision making could not be verified but for Terai floods the data provided by SERVIR strongly influenced resource allocations decisions by World Food Program (WPF). |
| *Ladcover mapping for Greenhouse Gas Emissions Inventory in Rwanda* | ● Surveys, focused group interviews | *Potential impacts identified.*<br>● A lot of factors hindered direct and maximum use of the product but potential impact points towards improvement in policies related to land cover changes and Greenhouse gas (GHG) emissions. |
| *Landcover mapping for GHG Emissions Inventory in Zambia* | ● Surveys, focused group interviews, semi-structured interviews | *Potential impacts identified.*<br>● Induce country level participation for Reducing emissions from deforestation and forest degradation (REDD) and help monitor plantation structures |
| *Rapid Response Mapping and Agricultural Monitoring for Food Security in Nepal* | ● Economic valuation model (limited information available) | *Actual impacts identified.*<br>● There were observed property damages and losses averted due to use of rapid response mapping in Nepal that provide streamflow information to guide irrigation planning. The agricultural monitoring service was observed to have helped in improved efficiency of service delivery by government ministries and other product users. |

## 2.2. *Takeaways from Case Studies*

The SERVIR audience or practitioners of Economic Impact Assessment, should reflect on the parallels between the types of economic impact questions that have been asked of other EO-related projects compared to those applied to SERVIR projects. Critical reflection of the key findings presented below will help pose similar questions to SERVIR services that are most relevant to stakeholders. A common misconception of EIA is the assumption that the final results must be expressed solely in monetary terms. Rather, both quantitative and qualitative methods can lead to a wide range of findings that may be articulated in more compelling or relevant terms. For a thorough discussion of the limitations and considerations of each study, readers are advised to refer to the full publications or reports.

The impact assessments (IAs) conducted by VALUABLES and SERVIR describe the impact of EO across diverse applications, from frost early warning systems to forest fire monitoring, flood early warning systems, and deforestation monitoring. This set of examples demonstrates the applicability of multiple types of EIAs, including willingness to pay models, VOI, DICE, DID, contingent valuation, measurement of loss avoided, among others, as well as qualitative approaches such as surveys, focus groups, and structured and semi-structured interviews.

The assessments of SERVIR services varied in their use of quantitative and qualitative data and interviews vs. modeled EIA approaches. The EIAs that used modeled approaches include the Bangladesh Flood Early Warning System (economic valuation model/stated preferences valuation/measurement of loss avoidance), the Frost Mapping, Monitoring, and Forecasting System (measurement of loss avoidance), and the Forest Fire Monitoring in Guatemala service (contingent valuation method) which were able to isolate impacts such as a savings of $73.6 million dollars with access to a 3-day flood warning, savings of $80.47 per Kenyan tea farmer (the equivalent of 25 days of household food spending, almost a full year of a child's school tuition, or a full year of household health spending), and that users of the fire monitoring tool would be willing to pay $78 per year for the service. We note that while these SERVIR examples used EIA to quantify impact in dollars, EIA can also be used to calculate other types of impact metrics beyond dollars saved.

The majority of the SERVIR assessments described in Table 1 rely on structured and semi-structured interviews with end-users and beneficiaries of the SERVIR services. These types of data are useful for providing narrative impact stories, personal experiences, and examples of tool use (including unintended uses). Currently, SERVIR has robust guidance and best practices for generating use cases and narrative-driven impact stories integrated into the service planning toolkit. However, these types of IAs alone may not provide the types of information that other approaches exploiting more robust datasets and economic models are able to achieve, such as number of lives saved, dollars saved, number of people impacted, loss avoided, acres insured, among many others. For example, the IA of the SERVIR service monitoring Harmful Algal Blooms in El Salvador relied on interviews with SERVIR service end-users and beneficiaries. The results of this IA included several examples of how the service had been used, as well as documentation of people's perception of the tool, but it did not provide the information necessary to quantify the number of people impacted, the dollar value of the impact the service has had (on health, tourism, fisheries, etc), the number of lives saved, or other potential metrics. VALUABLES evaluated a similar service using a discrete choice random utility maximization (RUM) approach that required very large datasets and more intensive economic modeling. This EIA found that the expected value of perfect information (EVPI) on HAB occurrence for recreational users of a subset of California lakes would be $2.76 million dollars. While narrative impact stories such as those generated for the SERVIR HAB service have great value and can be powerful communicators of service impact, some audiences, including potential donors, require more quantitative valuations of a service's impact.

Observations from this review of selected IAs include:
1) There is a tradeoff between IA approach, available resources and data, and the potential audience for the results. It may not be necessary or possible to use the most cutting-edge model or data and resource intensive approach to generate useful IA insights. However, the ultimate impact metrics are dependent on the available data and the implemented approach. EIAs are capable of generating meaningful metrics that complement narrative use cases and personal experiences of services.

2) Data intensive approaches to EIA benefit from incorporation of data collection into the service design and monitoring efforts from the implementation of the project, resulting in a more complete and extensive database of information to pull on for EIA. Retroactive data collection may be less complete and more resource intensive. For example, the report on the Bangladesh Flood Early Warning System was limited to one recent flood event, because data collection for EIA was not implemented until late in the service's use.

3) EIAs conducted by VALUABLES and SERVIR demonstrate the applicability of several approaches to conduct an EIA of EO data, however other approaches have also been applied with some success, and their potential application should also be considered in SERVIR's future EIA activities. For example, time series approaches have been useful in quantifying the impact of a malaria early warning system and volcanic ash advisories. However this requires access to data before and after implementation of the service and would therefore benefit from having more comprehensive EIA-focused data collection throughout the service life span, including a robust, pre-service implementation baseline.

We also direct readers to the NASA ESD primer on *Measuring the Socioeconomic Impact of Earth Observations* (NASA 2023), which provides a comprehensive list of EIA lessons learned for the broader EO community.

## 3. Survey of the SERVIR network

This section presents the results from a comprehensive survey conducted across SERVIR's hubs and services to identify questions that can be addressed in the form of deep dives for EIA. The primary objective of this survey was to generate exploratory ideas from the entire SERVIR network regarding services or tools that benefit from understanding their economic impact. Some conversations led to questions outside the realm of impact assessment and tended toward ideation of including socioeconomic datasets and expertise as part of services in future. We gathered insights from the hubs that implement SERVIR in Southeast Asia [implemented by the Asian Disaster Preparedness Centre (ADPC)], West Africa [implemented by the International Crops Research Institute for the Semi-Arid Tropics (ICRISAT)], Amazonia [implemented by the International Center for Tropical Agriculture (CIAT)], and Hindu-Kush Himalaya [implemented by the International Centre for Integrated Mountain Development (ICIMOD)]. We conducted strategic listening sessions with SERVIR Science Coordination Office (SCO) thematic leads and regional coordinators, as well as select SERVIR Applied Science Teams (ASTs) when suggested by the Hubs. USAID mission representatives were briefed and invited to provide additional input and feedback prior to the final synthesis (next section). The questions aligned with respective implementer/hubs have been segregated into the five thematic areas of SERVIR tools:

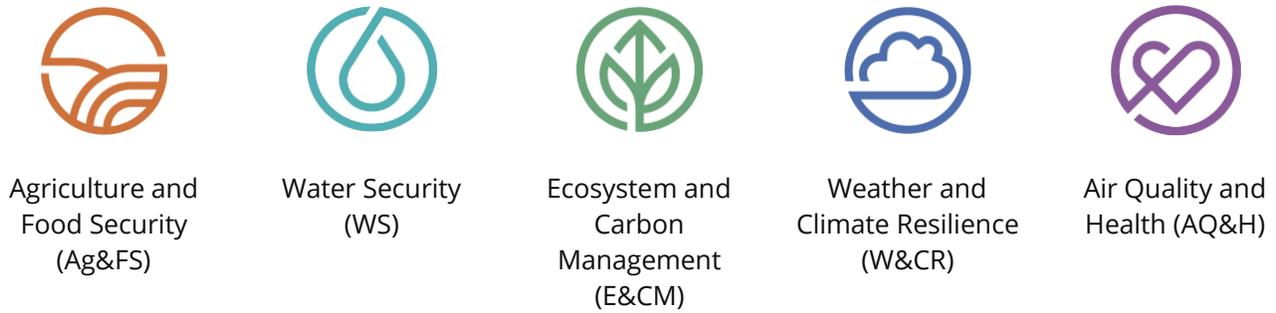

Agriculture and Food Security (Ag&FS)   Water Security (WS)   Ecosystem and Carbon Management (E&CM)   Weather and Climate Resilience (W&CR)   Air Quality and Health (AQ&H)

**Figure 2:** Thematic service areas of relevance to this study

The survey also includes potential EIA topics for SERVIR tools and techniques to be implemented for a future Central America Hub along with EIA topics regarding activities at the previous Eastern and Southern Africa Hub [implemented by the Regional Centre for Mapping of Resources for Development (RCMRD)]. The SERVIR tools and services related to the EIA topics identified in the survey are mentioned in the footnotes.

The SERVIR network is curious about EIA. More than 60 questions related to EIA came forward. Most respondents were motivated by understanding, measuring, and communicating the real implications of the services they are co-developing, but they lack the training, tools, and time to conduct EIA themselves. Most also referenced communications, MEL, and impact stories as starting points to brainstorm questions they thought EIA could help address. A few economists in the SERVIR network nudged conversations forward and encouraged the scientists and communicators to think about the quantitative and qualitative methods that economics can bring to bear. The tables from Table 2 to Table 7 represent the EIA questions of interest to the SERVIR network, organized by thematic and geographic area.

**Table 2:** Questions and issues identified for activities in Southeast Asia

|  | **Science Coordination Office** | **Hub** | **AST** |
|---|---|---|---|
| **Ag&FS** | 1. What is the impact of socioeconomic variables on crop yield?<br>2. How to incorporate and evaluate the impact of socio-economic variables on crop yield modeling (Myanmar).[1]<br>3. What is the economic impact of Coup on crop(rice)/food security (Myanmar)[1]<br>4. What is the economic impact of using drought nowcasts and forecasts at the regional level in the Lower Mekong Region through service developed using the Regional Hydro-Extreme Assessment System (RHEAS)? | 5. Impact of socioeconomic variables in crop yield (Myanmar).[1]<br><br>6. Impact of drought & crop yield forecasts on food import/export decision-making (RHEAS). How do these forecasts reduce exports? Lower interest in effects on imports..[6] |  |
| **WS** | 7. Where water is located vs. known rice crops? Did flood inundation maps help reduce crop losses?<br>8. Are the benefits of floodwaters calculable (irrigation, sediments, fertilizers)? |  |  |

|  | **Science Coordination Office** | **Hub** | **AST** |
|---|---|---|---|
| **E&CM** |  |  | 9. What are the economic impacts of deforestation mapping?[5] (Cambodia) |
| **W&CR** | 10. What human factors play a crucial role in recovering from disasters? What are the population's vulnerability to cope with different disasters?[3] | 11. What is the economic benefit of a successful flood and drought early warning system in agriculture in Myanmar?[1]<br>12. Reduction in fires / forest fire hotspots in hectares.[4] |  |
| **AQ&H** |  | 13. What is the economic value to the health impact (diseases caused by respiratory illness from poor air quality) due to the implementation of Air Quality Monitoring and Forecasting in either Thailand or Lao PDR?[4] |  |

[1] Geospatial Applications for Food Security and Sustainable Landscapes in Burma (Myanmar); [2] Supporting Flood Emergency Preparedness for Myanmar; [3] Enhancing Anticipatory Actions for Disaster and Climate Resilience; [4] Enabling Sustainable Landscape-Scale Agricultural Management through Fire and Air Quality Monitoring; [5] Building a Dashboard to Monitor, Evaluate and Report Landscape Improvements in Cambodia; [6] Enhancing Drought Resilience and Crop Yield Security for the Lower Mekong

**Table 3:** Questions and issues identified for activities in West Africa

|  | **SCO** | **Hub** |
|---|---|---|
| **Ag&FS** | 14. What is the economic impact on food security in Senegal as the hub applies the service which uses satellite data to map crop types and conditions in Senegal?[1]<br>15. P-Locust service provides real-time monitoring of Locust in WA countries. What is the economic impact of using the P-Locust tool as it helps the end-user in planning early for upcoming locust outbreaks?[2] | 16. Ag and meteorological services use forecasts to give advice to users to choose seed types that will be better for next season. Better harvests are expected if forecasts are good and match with seed types. What is the economic impact of forecasts? How do these forecasts and crop condition model payouts for insurance or extreme events?[1,6] |
| **WS** | 17. What is the impact of the monitoring ephemeral water bodies service (WENDOU) in Senegal as it helps pastoralists to understand their livestocks water accessibility?[3] | 18. Impact of WENDOU service on pastoralist livelihoods.[3] |
| **E&CM** | 19. Comparative analysis between forest incentive programs and carbon market payments[5]<br>   a. Adding the value of carbon as a new layer/feature (like socioeconomic data in crop yield models). How could adding monetary values to pixels prioritize land protection, and/or<br>   b. The national investments are a SERVIR Carbon Monitoring Pilot (S-CAP) sub-question. Aside from that, the REDD+ could be another way to look at this. (what is the value of this parcel of land, and can this help us figure out where to target who invests in this) | 20. What would be the impact of using the Galamsey Service [on forests, on water quality, and on livelihoods in legal mining sectors?]?[4]<br><br>21. What will be the actual cost to the environment of Charcoal production?[5]<br><br>22. How will information have an impact on the value chain? [Conceptualizing everything in terms of Carbon and the various services associated with it] |

|  | SCO | Hub |
|---|---|---|
| **W&CR** |  | 23. What is the socio-economic impact of floods on crop yield?[1] |

[1] Crop Type Mapping and Condition Assessment in Senegal; [2] Locust Monitoring; [3] Monitoring Ephemeral Water Bodies in Ferlo, Senegal; [4] Monitoring of Artisanal Mining (Galamsey) in Ghana; [5] Charcoal Production Site Monitoring Service for West Gonja and Sene Districts in Ghana; [6] Sub-seasonal to Seasonal Forecasting

**Table 4:** Questions and issues identified for activities in Amazonia

|  | SCO | Hub | AST |
|---|---|---|---|
| **E&CM** | 24. What is the economic impact of using zero deforestation maps on agricultural production of small-scale farmers[1] [ongoing work in this area – Forest Conservation Evaluation Tool (FCET) (Allen Blackman)]<br>25. Cost Benefit Analysis of monocrop cacao vs agroforestry (shaded) cacao. How do we tip the scale so that farmers practice agroforestry instead[1] [looks closer to Forest Conservation Targeting Tool (FCTT), opportunity cost layer?] | 26. Economic dependency on mangrove ecosystems[2] (Guyana)<br>27. How is The Radar Mining Monitoring tool (RAMI)[3] (gold mining) service used? What is the Cost? [Availability of information may be uncertain]<br>28. Is the service focusing on efficiency of mapping or monitoring agroforestry areas being incorporated by the Brazilian Agricultural Research Corporation (EMBRAPA) and INDRI? What is the opportunity cost of monocrop farming vs. cacao/agroforestry system).[1]<br>29. Comparative study of S2S adoption/use across hubs/regions[4]<br>30. What is the economic value of fire forecasts?[4] | 31. Efficiency of mapping/monitoring agroforestry areas[1] (where is cacao growing now?)<br>32. Opportunity cost question[1] (mono-crop farming vs cacao/agroforestry system)<br>33. Why is it that farmers are not interested in certification? If they get the products certified, they might be able to acquire loans for agricultural production.<br>34. What happens to the land after deforestation? How long does land go fallow?<br>35. How much more money would a farmer make if they were involved in shaded plantations?<br>36. Why has there been an observed decrease in deforestation over the years?<br>37. Why are farmers switching to monocrop cacao?<br>38. What is the impact of knowing where the farmers are located? - Is the area ecologically important? |

[1] Ecosystem Services Modeling in the Amazon's Forest-Agricultural Interface; [2] Monitoring and Evaluation of Mangroves in Guyana; [3] Monitoring of Gold Mining in the Brazilian Amazonia; [4] Forecasting Seasonal to Sub-Seasonal Fire and Agricultural Risk from Drought

**Table 5:** Questions and issues identified for activities in Hindu Kush Himalaya

|  | SCO | Hub |
|---|---|---|
| **Ag&FS** | 39. What is the economic impact of Agro-Met service used for Nepal and Bangladesh? This service provides baseline crop type maps at sub-district level. What are the important decisions made using this service and what's their economic impact?[1]<br>40. What is the economic impact of crop monitoring and assessment service as this service provides crop type and yield information at field-scale?[2]<br>41. What is the economic impact as end-users use the food security vulnerability information system of Nepal to make the decisions?[3]<br>42. What is the economic impact of regional drought monitoring and outlook systems for South Asia?[4] | 43. What is the impact of using early drought forecasts/ drought monitoring service on agriculture? What are the economic indicators we can develop to track the impacts?[2,4] (Nepal) |

|  | SCO | Hub |
|---|---|---|
| W&CR | 44. What is the economic impact of using flood inundation maps on agriculture?[2,5] (Nepal) | 45. What is the economic impact of using flash flood forecasts [on lives, livelihoods, or certain sectors such as agriculture]?[6] (Nepal) |

[1] Agro-met Advisory Service for National/Local level Planning in Nepal and Bangladesh
[2] Crop Monitoring and Assessment
[3] Food Security Vulnerability Information System of Nepal
[4] Regional Drought Monitoring and Outlook System for South Asia
[5] Flood Inundation and Flood Occurrence Mapping
[6] Enhancing Flood Early Warning Services in the Hindu Kush Himalaya Region

**Table 6:** Questions developed for activities in Eastern & Southern Africa

|  | SCO |
|---|---|
| Ag&FS | 46. Kenya food balance sheet – what is the value of EO based crop modeling or decision making at country level?[1]<br>47. What is the value of RHEAS on digital food balance?[1]<br>48. Impact of Index-based livestock insurance [1,2]<br>49. What is the [lasting] impact of frost monitoring and forecasting service [on tea producers]?[3] |
| E&CM | 50. What is the economic impact of carbon monitoring [to national governments, or to community forest managers]?[4] |
| W&CR | 51. What is the value of VIA (Vulnerability Impact Assessment) analysis on climate adaptation/resilience to climate?[5] |
| AQ&H | 52. What is the economic impact of Air Quality Monitoring on health? |

[1] Regional Cropland Assessment and Monitoring Service; [2] Kenya Rangeland Assessment and Monitoring Service; [3] Frost Monitoring and Forecasting Service;
[4] Land Use Land Cover and Change Mapping Service for Eastern and Southern Africa; [5] Climate Change Vulnerability, Impacts, and Assessments Service

**Table 7:** Questions developed for activities in Central America*

|  | SCO |
|---|---|
| WS | 53. What is the value of EO products to stem hurricane-induced flood impacts on agriculture (e.g., hurricanes Eta and Iota)<br>54. What are potential economic impacts of integrating GEOGLOWS into Belize's early warning system for floods? (Currently Belize manually checks many but not all gauges twice a day.)<br>55. Follow-up with participants of GEOGLOWS training to see if any ideas come out of their projects?<br>56. What are some lasting impacts in El Salvador: The Government of El Salvador estimated that the country saved funds due to the use of NASA MODIS data for monitoring red tides, both due to unsafe fish that were not harvested, and people who did not get sick from eating unsafe marine products. |
| E&CM | 57. What are the region's forest carbon stocks theoretically worth?<br>58. Can we use economics to justify forest conservation, or are the opportunity costs for converting biodiversity-rich forests into other land uses higher?<br>59. Could observing the above question be a value add that SERVIR could provide to the region, in terms of an improved understanding of the region's natural capital? For instance, and in parallel, in 2009 World Resources Institute (WRI) provided the Government of Belize with a valuation of the country's coral reefs, and that helped to provide key understanding about how much the nation's economy is tied to tourism and fisheries related to the reef.<br>60. S-CAP application - looking at the value of carbon sequestered by a small, protected area. Potential of this kind of model of co-management (between community and National Protected Areas System) for carbon sequestration.<br>61. "Because of SERVIR's support for the development of SERVIR-Mesoamerica—a pilot project for the Geospatial Information System for Fire Management (SIGMA-I), the Government of Guatemala estimates that *x* amount of funds was saved due to the prevention of forest fires." |

| | SCO |
|---|---|
| **W&CR** | 62. Identification of precarious settlements → simulate how knowing where precarity exists can facilitate government agencies' preparation for disasters and/or distribution of resources pre/post disaster. |

*Many of these questions are inspired by proto-services in Central America, https://servirglobal.net/where-we-work/central-america

## 4. Readiness assessment and analysis of gaps and opportunities

Conversations with a broad set of groups throughout the SERVIR network led to a series of economic impact questions or hypotheses that can be considered for EIA. USAID partners (e.g., agreement officer representatives, climate advisors, and environment officers in the immediate SERVIR network) also provided additional input and feedback. Criteria for selection include the importance of each question and the feasibility of conducting the EIA. The importance and feasibility criteria are defined as follows:

- **Importance:** relevance to hub, USAID, NASA, stakeholders, and other interested parties; alignment with SERVIR's scope (funding, strategy / theory of change); whether SERVIR is co-developing similar services in multiple regions (sets us up to replicate similar assessments in multiple regions).
- **Feasibility:** access to and availability of users and co-developers; availability of socio-economic impact data; methods exist in the economic literature to conduct the assessment; answerable in a reasonable time frame with available resources; maturity of the service.
- **Schedule (phase of service):** A well-established service might have better socio-economic impact data as opposed to a service that is just delivered or is still in the development phase. On the other hand, newer services could be designed with EIA in mind from the outset, eventually making these more feasible to assess in the future.

This section presents a detailed overview of economic analyses of potential interest to SERVIR across all thematic service areas (Table 8: Agriculture and Food Security, Table 9: Water Security, Table 10: Ecosystem and Carbon Management, Table 11: Weather and Climate Resilience, Table 12: Air Quality and Health and Table 13: Cross-Hub Regions). We condense the 60+ questions from the previous section into roughly half, combining similar questions and prioritizing those that focus on impact assessment (as opposed to socioeconomic data in general). Each service area is examined through these specific questions and the potential methods and expected findings that could be discovered if pursued. The analyses of importance, feasibility, and schedules, aim to help the SERVIR network understand how economic impacts of different interventions, services, and initiatives can be assessed in their respective sectors.

This section also emphasizes the potential role of these EIAs – or at least the process of asking and refining these types of questions – in improving the design and evaluation of SERVIR services. From there, we aim to inform better decisions, improved practices, and enhanced resilience across different regions, thanks to the skillful application of EO-driven services and the capacity built along the way.

### 4.1. *Assessment for Agriculture & Food Security*

Changes in climate have profound effects on agriculture and food security, posing risks like droughts, floods, and pests, which can severely impact agricultural production and rural livelihoods. Conversely, food production contributes to climate change through greenhouse gas emissions and environmental degradation like deforestation and water pollution. Balancing food provision for growing populations with climate resilience is a global challenge. SERVIR addresses these challenges by collaborating with agricultural decision-makers at various levels, supporting enhanced monitoring, forecasting, and early warning systems for agriculture. Using Earth observations, SERVIR assists in producing crop maps crucial for national agricultural monitoring and greenhouse gas reporting. It also aids in rangeland monitoring for pastoralists, leveraging satellite data for vegetation and precipitation. Predictions of temperature and precipitation facilitate informed decisions for farmers to boost yield and income, while crop yield forecasts inform policymaking and aid distribution during crises like droughts.

**Table 8:** Refined questions related to Agriculture & Food Security, and the related potential EIA methods, expected findings, and brief descriptions of feasibility, importance, and schedule

| Questions Posed by SERVIR | Potential Methods | Expected Findings | Feasibility | Importance | Service Schedule |
|---|---|---|---|---|---|
| 1. What is the impact of socio-economic variables on crop yield, or how to incorporate socio-economic variables on crop yield modeling, and what is the impact (Myanmar)? | Machine learning models, econometric models with machine learning applications, simulation models. | Socio-economic factors like household income, education, gender, land size, government support policies, and technology significantly impact crop yield. | Data uncertain, but economic data from the statistical department may be used. | High | In Design |
| 2. What is the economic impact of the coup on crop (rice) production and food security (Myanmar)? | General equilibrium models, macroeconomic models, simulation models. | Coup will adversely affect the economy in various ways. | Low data availability, but statistical department data can be used | High | In Design |
| 3. What is the economic impact of projections of agricultural and meteorological services? How do these forecasts and crop condition model payouts for insurance or extreme events? | Social cost benefit analysis, econometric models, VOI game theory, probabilistic risk models, DID model | Insurance aids beneficiaries: affordable insurance promotes higher-value crops. | Low without Eastern & Southern Africa (ESA) hub connections to insurance companies | Moderate to High | Delivered |
| 4. What is the impact of early drought forecasts/ drought monitoring service? What are the indicators we can develop to track the impacts? (Nepal) | Farm level surveys, Dynamic modeling, VOI | Early warning on drought positively impacted farmers or decision-makers | High, as the hub generates drought monitoring and seasonal outlooks | Moderate | January 2019 - present |
| 5. What is the economic impact of RHEAS in lower Mekong on drought resilience and crop yield security? | Yield modeling and regression analysis, risk assessment modeling | Improved planning by farmers and policymakers based on RHEAS predictions | High, hub may have relevant data | High | Delivered |
| 6. What is the economic impact on food security in Senegal or Ghana as the hub provides crop type | Input-output models, partial equilibrium models, econometric models | Improved planting and harvesting decisions leading to increased yields | Low-Moderate, access to extension agents and farmers may | Moderate to High | In Development |

| Questions Posed by SERVIR | Potential Methods | Expected Findings | Feasibility | Importance | Service Schedule |
|---|---|---|---|---|---|
| maps and conditions? | | | be challenging | | |
| 7. What is the economic impact of using the P-Locust tool as it helps the end-user plan early for upcoming locust outbursts? | Simulation models, VOI, Spatial Analysis Tools | Informed pest control measures and increased cropland production | High, hub has connections with on-the-ground pest management activities | Moderate | In Development |
| 8. What is the economic impact of crop monitoring and assessment service as this service provides crop type and yield information at field-scale? | General equilibrium models, econometric models, dynamic simulation models | Improved farm management decisions leading to increased productivity | Low, service is still in development | High | In Development |
| 9. What is the impact of frost monitoring and forecasting services? | Stated preference methods, computable general equilibrium models, econometric models | Significant cost savings for tea farmers in Kenya | Low, no active hub in E&S Africa as of 2023 | Low | Ended in 2021 |
| 10. Cost Benefit Analysis of monocrop cacao vs agroforestry (shaded) cacao. How do we incentivize farmers to practice agroforestry? | Cost-benefit analysis, net present value, farm level surveys, and interviews | Improved productivity and commercial value of crops with informed farm management decisions | Low, no data available | Moderate | In Development |

### 4.2. *Assessment for Water Security*

Water security is vital for poverty eradication, gender equality, food security, and ecosystem preservation, but climate change and population growth intensify water-related stresses, jeopardizing water availability for irrigation and domestic use. Competing demands for freshwater within and between nations pose challenges for water security and diplomacy, particularly affecting vulnerable groups. SERVIR supports the enhancement of sustainable freshwater access and fosters cooperation through Earth observations-driven services and capacity building. Earth observation data, including hydrologic models, monitor and forecast water availability and quality, aiding in decision-making for human consumption, irrigation, and reservoir management. Satellite-based estimates track surface water extent, assisting pastoralists and monitoring international watersheds like the Mekong River. Long-term trends guide water project planning and allocation, while monitoring surface water quality informs management strategies against pollutants and algal blooms. Integrated satellite and ground measurements assess aquifer status and groundwater dynamics.

**Table 9:** Refined questions related to Water Security, and the related potential EIA methods, expected findings, and brief descriptions of feasibility, importance, and schedule

| Questions Posed by SERVIR | Potential Methods | Expected Findings | Feasibility | Importance | Service Schedule |
|---|---|---|---|---|---|
| 11. What is the economic impact of Wendou Service on pastoralist livelihoods? | Econometric Models, Time Series Analysis, Social Accounting Matrix | Economic impact analysis can reveal the service's contribution to economic and livelihood stability, reduced livestock losses, and overall economic resilience | Moderately feasible, contingent on data availability for EIA | High, as the service significantly affects the livelihoods of pastoralists | In the development phase |
| 12. What is the value of EO products to stem hurricane-induced flood impacts on agriculture (e.g., Eta/Iota)? | Agent-Based Models, Risk Assessment Models, Computational General Equilibrium Models, Machine Learning Models, Time Series Analysis | Implementation of early action protocols (EAP) using EO data could have mitigated crop losses. Specific EO products like GEOGLOWS ECMWF Streamflow prediction service and HYDRAFloods could inform EAPs | Low feasibility as of 2024 due to the absence of a hub | High, particularly relevant to USAID Central America & Mexico (CAM) Mission and the President's Emergency Plan for Adaptation and Resilience (PREPARE) | Not currently a service in Central America |

### 4.3. *Assessment for Ecosystem & Carbon Management*

Healthy ecosystems, including forests and landscapes, provide essential benefits such as food, clean water, biodiversity, and climate regulation by managing carbon. However, deforestation, unsustainable land management, and land degradation globally jeopardize these ecosystems and contribute to greenhouse gas emissions. SERVIR collaborates with partners, including Indigenous communities, to bolster institutional capacity for sustainable resource management, utilizing Earth observations to address climate change mitigation and adaptation. To support sustainable ecosystem management and carbon monitoring, SERVIR develops tailored Earth observations-driven services. These services enhance land cover monitoring, invasive species mapping, forest vulnerability assessment, and management strategies. Additionally, SERVIR assists with issues like fire detection, biodiversity conservation, and carbon reporting, offering near-real-time deforestation and fire alert systems. The network aids countries in implementing national Monitoring, Reporting, and Verification (MRV) systems for REDD+ initiatives, supplying data for payments and ecological forecasting. Furthermore, SERVIR's services cater to the growing demand for certification and access to deforestation-free products.

**Table 10:** Refined questions related to Ecosystem & Carbon Management, and the related potential EIA methods, expected findings, and brief descriptions of feasibility, importance, and schedule

| Questions Posed by SERVIR | Potential Methods | Expected Findings | Feasibility | Importance | Service Schedule |
|---|---|---|---|---|---|
| 13. What is the economic impact of Deforestation Mapping in Cambodia? | Spatial Econometric Models, Computable General Equilibrium Models, CBA, IOA | Government Revenues due to avoided deforestation, Greenhouse Gas emissions avoided due to alerts, cost savings, improved agricultural productivity, better land management and land use, sustainable biodiversity and reduced risks to hazards related to landslides | High. But determining beneficiaries will take time. | Moderate | Delivered |
| 14. What will be the actual cost to the environment for charcoal production? | Environmental Valuation, Carbon Accounting, Life cycle assessment, LCA, IOA, Externality Analysis | Government Revenue vs. Environmental Cost, emission levels due to deforestation and degradations, excursion violation and adherence. | Moderate. Data in revenue required | Moderate | In Process |
| 15. What is the economic impact of Galamsey in terms of environmental degradation, greenhouse gas emissions, loss biodiversity, declines in water quality? How is RAMI service used? What is the cost/benefit of implementing the service? | Environmental Valuation, Social Cost Benefit Analysis | Operational Cost Savings to relevant ministries. | Moderate due to revenue data and funding gap | High | Galamsey Service: In Development; RAMI Service: Delivered |
| 16. What is the economic impact of using zero deforestation maps on agricultural production of small-scale farmers? | CBA, SCBA, IOA, Spatial Econometric Models | Reduced expansion of small-scale agriculture in the pilot area. | Difficult in Assessing Impact | High | In Development |
| 17. What are the costs and benefits associated with implementing the mangrove monitoring ecosystem in Amazonia? What are the economic benefits provided by mangrove ecosystems, and what are the associated environmental costs of removing and degrading mangrove ecosystems? | Environmental Valuation, Carbon Accounting, Life cycle assessment, LCA, IOA, Externality | Opportunity Cost of not clearing mangrove forests and leaving them intact in terms of flood protection and carbon sequestration. | Moderate to High | Moderate to High | In Development |
| 18. What is the effectiveness of mapping/monitoring agroforestry areas (where is cacao growing now?) | Spatial Econometric Models, CBA, SCBA, IOA | Cost Savings | Low to medium due to funding gap | Moderate to high | In development |

| Questions Posed by SERVIR | Potential Methods | Expected Findings | Feasibility | Importance | Service Schedule |
|---|---|---|---|---|---|
| 19. What is the economic value of fire detection in terms of improvements in forest management in Amazonia and Nepal? | CBA, IOA, SCBA, Natural Resources Accounting | Better forest management and emissions reduction | Low. Need to understand end users | Moderate | Delivered |

### 4.4. *Assessment for Weather & Climate Resilience*

Climate change is leading to more frequent and intense hazardous weather events, highlighting the importance of early warning systems in minimizing casualties and economic losses. However, global and local observing systems require strengthening to provide timely and accurate forecasts that prompt effective action. SERVIR focuses on co-creating tools and enhancing capacity to integrate forecast information into decision-making at local, national, and regional levels. Collaboration with local expertise ensures satellite data's relevance and usability. Working with host country partners, SERVIR improves forecast accuracy and lead time by analyzing locally and remotely collected weather and climate data. Services include temperature and precipitation forecasts, specialized hazard forecasts, and long-range climate outlooks. Examples range from landslide risk forecasts to drought advisories and flash flood warnings. Disseminating this information to various end-users supports early action and decision-making, safeguarding lives and livelihoods.

**Table 11:** Refined questions related to Weather & Climate Resilience, and the related potential EIA methods, expected findings, and brief descriptions of feasibility, importance, and schedule

| Questions Posed by SERVIR | Potential Methods | Expected Findings | Feasibility | Importance | Service Schedule |
|---|---|---|---|---|---|
| 20. What human factors play a crucial role in recovering from disasters? What is the population's vulnerability and ability/capacity to cope with different sorts of disasters? [Application - South East Asia (SEA)] | Surveys, Cost-Benefit Analysis, Social Returns on Investment Analysis (SROI), Participatory Action Research (PAR). | Identifying areas at greatest risk to suffering impacts from hazard events. Areas with greater risk at great impact cost more to recover from. | Moderate. This is a very broad query and could use refinement as it can encompass sociologic and economic aspects. Feasibility depends on available population, demographics and infrastructure information needed to assess vulnerability. A starting point is the Mekong X-Ray tool which contains population data, and other infrastructure information from. More detailed understanding and impacts would require field information we do not have at this point. A possible collaboration could be with the World Food Programme and Food and Agriculture Organization of the United Nations (FAO)-Asia who are active in Anticipatory Action activities. | Moderate. Targeting areas at greater risk to suffer from the hazard impacts is important, however, necessary detailed population and infrastructure information is typically difficult to obtain and keep up to date. | Mekong X-Ray is currently under development with plans to expand. Data is already integrated in MRC's Flash Flood Guidance System to generate a flash flood index and estimate hospitals, people etc. affected. |

| Questions Posed by SERVIR | Potential Methods | Expected Findings | Feasibility | Importance | Service Schedule |
|---|---|---|---|---|---|
| 21. Where water is vs. where we know rice agriculture lies? Did flood inundation maps help reduce losses? | Deterministic forecasts, Household surveys and Value of Information(VOI) | Flood inundation maps may or may not reduce losses depending upon their extent and region of uses. Efficient use of early warning systems will result in higher economic benefits pertaining to agriculture. There could be reduced costs for monitoring and damage assessment. There is a potential to inform early actions to stem agricultural losses (e.g., early harvesting) | High/Moderate. Hindu Kush Himalaya (HKH) hub has focused on post-event damage assessments more than early or preventative actions related to rice crops. In SEA, AST Project tested a simplest model for this, looking at inundated areas and rice crops using an existing approach. Validating this approach may be challenging. This is a potential area where the World Food Programme/ FAO could be engaged, but no existing hub data or activities support this effort. Doesn't yield early action, just estimates damages. | High/Moderate | Some components have been delivered in both HKH and SEA; newer remote sensing methods continue to be developed in both regions. |
| 22. What is the Economic Benefit of a successful flood and drought early warning system in agriculture in Myanmar? | Randomized Control Trial, Cost-Benefit Analysis, Random Utility Models. | Preventive measures can be taken, and informed decisions on crop types and crop patterns will benefit the agricultural sector. | Low. Data uncertain. Data from the ground or end-user data may not be available at the local scale. Ex ante study now and ex-post study in 2025 when actual data is available. | High/Moderate. Hub recommended. | In Design |
| 23. What is the impact of frost monitoring and forecasting services? | Input-Output Analysis, Cost-Benefit Analysis, Surveys and Interviews, Scenario Analysis | Income saved in terms related to household food spending or school tuition | Low. There is not an active hub in ESA as of 2024. | Low | Ended in 2021 |

### 4.5. *Assessment for Air Quality & Health*

Clean air is crucial for human health, environmental well-being, and economic prosperity. However, globally, poor air quality contributes to millions of premature deaths annually. Industrialization, vehicle emissions, agricultural practices, and forest fires are major contributors to air pollution. Climate change exacerbates the situation by increasing the frequency and intensity of heat waves and droughts, which can worsen air quality and public health. SERVIR utilizes Earth observations to help decision-makers map sources of air pollution, track its movement across borders, and assess its impact on public health. These observations also aid in measuring rainfall and air temperature, improving forecasts to provide early warnings of health threats. Collaborating with regional and local partners, SERVIR supports air quality monitoring and public health services, informing actions to mitigate negative health impacts, such as issuing air quality warnings and implementing emission reduction policies. Furthermore, satellite data and forecasts can be utilized to enhance overall public health beyond air quality monitoring. This includes improved forecasts of vector-borne diseases like malaria, aiding local control programs in preparedness and response to disease outbreaks.

**Table 12:** Refined questions related to Air Quality & Health, and the related potential EIA methods, expected findings, and brief descriptions of feasibility, importance, and schedule

| Questions Posed by SERVIR | Potential Methods | Expected Findings | Feasibility | Importance | Service Schedule |
|---|---|---|---|---|---|
| 24. What is the economic value to the health impact because of the use of Fire Risk mobile applications on Air Quality and Health in either Thailand or Lao PDR? | Randomized Control Trial, Difference-in-Difference Approach, Value of Statistical Life, Simulation Models | Implementation of Air Quality monitoring will help inform public health decisions and can also improve public health. It will also help the tourism department of Lao PDR | Moderate/High. Ex ante study now and ex post study in 2025 when actual data is available | Very High. Hub recommended and of great interest to USAID and regional partners. | In-Design |
| 25. What is the economic impact of using the SmokeWatch Tool for fire and smoke detection in Thailand? | Cost-benefit analysis, Input-output analysis, Econometric Analysis, Surveys and Interviews | Reduced fire, improved protection of agricultural assets, health and tourism and reduced insurance claims for property damages | High | Very High. Hub recommended and of great interest to USAID and regional partners | Delivered |

## 4.6. *Assessment for cross-hub region impact questions*

Several questions came forth relating to the potential for cross-region assessments that could determine comparative results of the same or similar information services in different geographic and development contexts. Such EIAs would require more coordination but could generate additional findings and foster cross-network knowledge exchange and learning.

**Table 13:** Refined questions pertaining to multiple hubs, and the related potential EIA methods, expected findings, and brief descriptions of feasibility, importance, and schedule

| Questions Posed by SERVIR | Potential Methods | Expected Findings | Feasibility | Importance | Service Schedule |
|---|---|---|---|---|---|
| 26. Comparative study of S2S adoption/use across hubs/regions | This might not be as much of an economic question but could take on more of an organizational capacity assessment, or sector-focused capacity assessments. | A comparison of capabilities and regular use of S2S by users across sectors or application areas like fire, flood/water security, drought, El Niño | Medium. Given the sprawling nature of this question, data could be a concern. If multiple regions are selected, local-level analysis could require years. It could be possible to ascertain differing impacts at regional/national levels and work with users to describe downstream benefits from their perspectives. | High. Information provided from S2S forecasts works toward PREPARE and EW4All goals to reach millions. | varied across SERVIR hub region |

| Questions Posed by SERVIR | Potential Methods | Expected Findings | Feasibility | Importance | Service Schedule |
|---|---|---|---|---|---|
| 27. Comparative analysis between forest incentive programs and carbon market payments / Would countries in SERVIR focus regions benefit more from participation in carbon markets or forest incentive programs? | Select one representative country per focus region. Compile information on the existence of forest incentive programs, and participation in carbon market programs (e.g., Voluntary Carbon Markets, World Bank's Forest Carbon Partnership Facility). Determine actual and potential economic revenues provided through both streams of programs. | In country x, local stakeholders / institutions can leverage x% more financing from forest incentive programs than from existing payments from the Forest Carbon Partnership Facility (FCPF). Alternatively, one finding could be the conclusion that both programs are not mutually exclusive in terms of participation. | High, This COmprehensive Approach will require high effort. | High | Not related to an existing service (S-CAP does not explicitly focus on carbon markets) |
| 28. What is the [differential?] economic benefits of GEOGloWS ECMWF Streamflow Prediction Services throughout multiple sectors across SERVIR regions (e.g., irrigation and/or flood early warning in Ecuador, flood forecasting in Brazil, flood early action in Nepal, community-based flood early warning in Malawi, reservoir operations in Honduras)? | VOI, Cost Benefit Analysis | Flood and streamflow predictions, if disseminated appropriately amongst the end users and if the information is utilized accordingly, will limit the losses and damages pertaining lives of people, houses etc. and on agricultural or water-based productions and incomes. | High. Several use cases and impact stories already exist, and connections to users / co-developers is very straightforward. There will be a breadth vs. depth question given the nature of these services. | Highly relevant to NASA, NOAA, USAID and USGS based on previous investments and expected strategy going forward; addresses both climate resilience and water security development outcomes | delivered and operational in HKH, Amazonia; in development in East Africa; information is accessible globally but not necessarily regionally- or locally tailored everywhere. |

### 4.7. *Overall Analysis*

The potential methods for economic impact analysis vary depending on the specific questions related to agriculture and food security, water security, ecosystem and carbon management, air quality and health, and weather and climate resilience. In summary, selecting the appropriate method depends on the specific research question, data availability, feasibility, and importance of the analysis. The valuation of a product, or in this case the information services derived from EO, depends significantly on its characteristics, including whether it is excludable or non-excludable and its use value and non-use value. Excludable goods are often valued based on their direct utility and the market demand they generate. Non-excludable goods may require more complex valuation methods, considering factors such as public benefit, externalities, and non-market values. Understanding the mix of use value and non-use value is crucial for accurately assessing the total economic value of a product or service to society. Investing in robust economic impact assessments can provide valuable insights for policymakers, stakeholders, and development initiatives in environmental and climate sectors.

In our analysis, the preliminary endeavors or projects concerning economic impact assessment are denoted as "deep dives." These deep dives are predicated upon the significance and feasibility of the inquiries garnered from interviews and listening sessions conducted with key stakeholders. Subsequently, a comprehensive table below was formulated encompassing all inquiries derived from these interviews, thereby furnishing a foundational resource to inform the progression of our study.

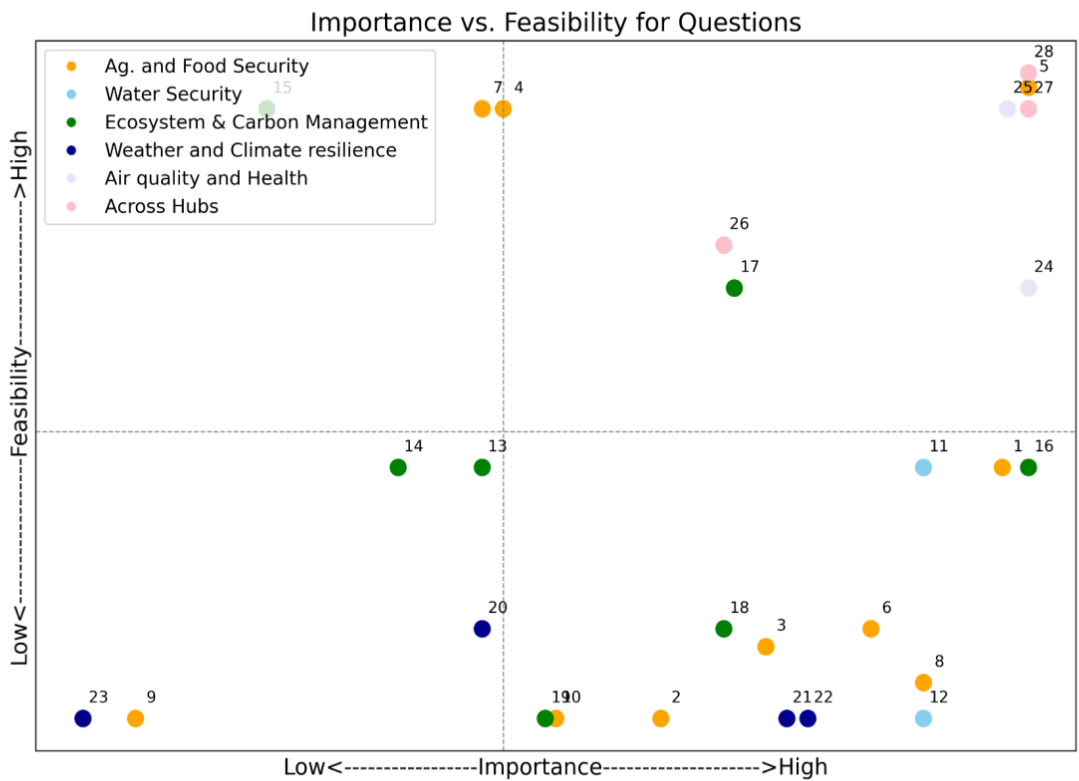

**Figure 3:** This graph illustrates the relative importance and feasibility of addressing questions related to five thematic areas. The x-axis represents feasibility, while the y-axis represents importance, with higher values indicating greater importance or feasibility. Each numbered dot point represents a question also listed in Tables above. Each theme is distinguished by a unique color.

The graph represents the importance and feasibility for the set of questions related to the five themes: Agriculture and Food Security, Water Security, Ecosystem and Carbon Management, Weather and Climate Resilience, Air Quality and Health, and General/Across Hubs, as referenced in the above tables. For each question, the importance and feasibility are depicted by two adjacent bars. The x-axis represents the feasibility of addressing each question and the y-axis represents importance for the same, ranging from low to high. Different colors are used to distinguish between the themes. Each theme is assigned a unique color scheme for better visualization. Numbers are used to highlight specific questions (25, 26, and 28). Custom legends are also provided to explain the color codes, representing different themes. Questions 25 [Air Quality and Health] and 28 [Across Hubs] exhibit the utmost significance and viability, stemming from the demands of Hubs and the availability of pertinent data.

Questions 25 and 26 [Ag. and Food Security] have been designated as the inaugural subjects for Deep Dive Projects, predicated on expressed interest from Hubs, SCO, NASA HQ, and USAID. Other factors of programmatic balance (i.e., diverse regional and thematic coverage) also motivated the downselection. The questions identified as high-importance and high-feasibility that were not selected can be considered "low-hanging fruits" that the SCO and/or hubs could pursue in the near term. Meanwhile, Question 28 constitutes an independent endeavor, conducted in partnership with NOAA, focusing on EIA pertaining to GEOGLOWS. For future EIA endeavors the top right quadrant in the graph is of utmost significance with higher importance and feasibility to pursue the assessment.

## 5. Conclusions and next steps

Based on prior performance evaluations of SERVIR, research by VALUABLES, and comprehensive interviews and discussions with the entire SERVIR network, we firmly establish that conducting EIAs holds paramount significance in the valuation process of EO data. First, EIA can play a crucial role in facilitating informed decision-making processes. In SERVIR's case, this is primarily through better-evaluated and better-designed services. By providing valuable insights into the economic benefits and potential risks associated with using EO data, stakeholders are empowered to make well-informed choices. Second, EIA can aid in more efficient allocation of resources toward EO programs. Understanding the economic ramifications ensures that resources are distributed effectively, maximizing program effectiveness and ensuring optimal fund utilization. Third, this analysis informs programmatic development endeavors by pinpointing areas where EO data can contribute significantly to economic growth, sustainability, and resilience. Fourth, EIA enables better risk management strategies by allowing stakeholders to anticipate and mitigate potential economic risks associated with EO data availability or quality disruptions.

EIA can be a cornerstone of the strategic planning and development of applied EO initiatives. The methods covered in the study identify a diverse range of EIA-related analytical approaches that have been applied elsewhere and could be employed more regularly by the SERVIR network. These include decision modeling, assessing the value of information (VOI), employing difference-in-difference (DID) models, estimating the value of statistical life (VSL), utilizing linear regression models, and calculating the net present value (NPV) of programs and services. These case studies also applied integrated assessment models, employing discrete choice random utility maximization models and revealed preference models. Surveys were also conducted to gather relevant data, and a multi-model framework and Bayesian updating processes were employed for analysis. These studies also incorporated structural listening sessions with stakeholders. Graphical representation and simulations were carried out using Bayesian Nets (BN). Causal mediation analysis and structured expert judgment (SEJ) exercises were also undertaken. Furthermore, classical model decision-making processes were incorporated. Finally, these studies integrated the Dynamic Integrated Climate Economy (DICE) Model and assessed real option value (ROV). SERVIR can triangulate the economic impacts of its services through a rigorous mixed-methods approach

## 5.1. *Selection of deep dive EIAs*

After conducting thorough research, interviews, and surveys, our investigations have led us to focus on two specific applied research inquiries or "deep dives" that emerged from our discussions.

1. The first area of focus pertains to evaluating the economic effects of implementing the Smoke Watch tool in Chiang Rai, Thailand. This tool utilizes data from the SERVIR air quality monitoring service. By delving deeper into this subject together with the SERVIR SEA hub, we aim to understand how adopting the Smoke Watch tool influences various economic factors in the region, such as health care and land cover changes. The project entails "needs assessment" and "data collection" for the analysis (Figure 4).
2. The second deep dive involves examining the impact of Subseasonal to Seasonal Forecasting on agricultural practices in Ghana. This entails analyzing how such forecasting techniques affect agricultural productivity, resource management, and resilience to climate variability in the region. Through this investigation, we seek to uncover insights that can inform agricultural strategies and policies in Ghana and similar contexts.

An ongoing collaboration with NOAA involves conducting an EIA focusing on the application of GEOGLOWS ECMWF Streamflow Prediction Services in Ecuador, which will serve as a third area of focus for EIA. We are currently developing the project description and technical details for the project report, conducting a feasibility assessment, collecting necessary data, and determining to what extent other countries can be included in the analysis.

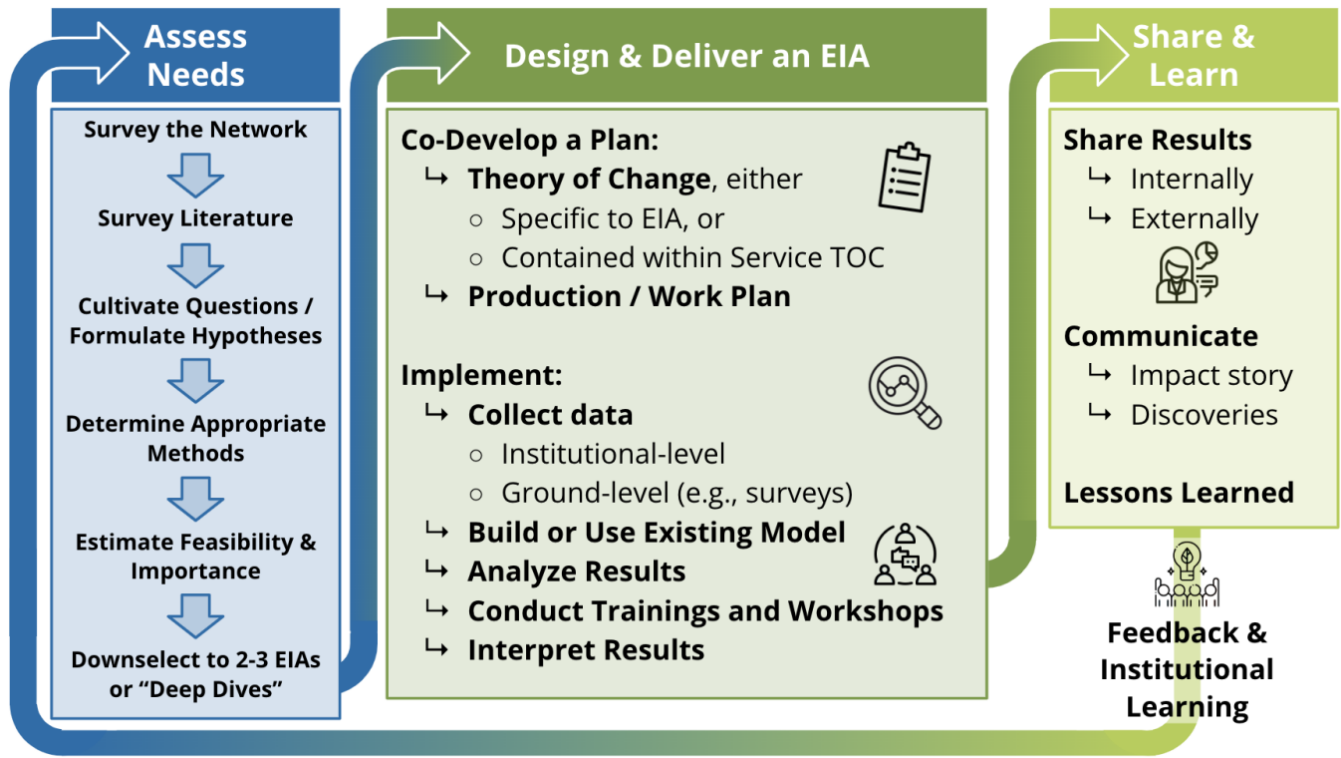

**Figure 4** Flow Chart on general approach to assessing needs for, conducting, and disseminating Economic Impact Assessment in the SERVIR context. This paper summarizes the "Assess Needs" phase, and each deep dive will follow steps identified in the "Design & Deliver" box. SERVIR will also share, disseminate, and learn from the findings.

## 5.2. *Further developing capacity in EIA and contributing to a growing community*

Beyond a series of intense and focused deep dives, it will be important to continue to dialogue, practice, and reflect on EIA's place in connecting EO to decisions and actions related to climate resilience and societal outcomes. To maintain transparency, accountability, and relevance of EIA to the SERVIR network and other interested parties, we will provide regular EIA updates. This will ensure that stakeholders within the network are consistently informed about the economic benefits and implications of EO initiatives. By keeping the SERVIR network abreast of the latest findings and insights from the EIAs, policymakers and practitioners can make well-informed decisions for allocating resources, formulating policies, and implementing projects. We expect each thematic service area personnel will apply different lenses and refine their theme-specific demands for EIAs. To the extent possible with hubs, we expect to co-develop service area-specific roadmaps for more regular incorporation of economic impact into services going forward. We will also strengthen ties with other related efforts, such as the CONVEI initiative. This is essential for fostering collaboration, knowledge-sharing, and synergies. Such collaborations do not only enhance the quality and credibility of EIAs but also facilitate the exchange of data, methodologies, lessons learned, expertise, resources, and best practices, ultimately benefiting the broader EO community. Finally, we will encourage hubs within the SERVIR network to request or offer impact assessments of their services to users and stakeholders. By incorporating impact assessments as an additional aspect of the services offered, hubs can better demonstrate the tangible benefits and value of EO applications to their communities. This proactive and participatory approach not only enhances user engagement and satisfaction but also enables hubs to tailor their services to meet their stakeholders' specific needs and priorities. Moreover, if users perceive value in these assessments, it can increase demand for EO services and foster long-term sustainability and impact.

## 6. Acknowledgements

The development of this paper would not have been possible without the support of the entire SERVIR network, the NASA Earth Action Program, and the USAID Bureau for Resilience, Environment and Food Security. We thank the following people and organizations from within and outside of the SERVIR network for their contribution to this phase of our study and the immediate report.

For programmatic guidance, oversight, funding, and supervision:
- Nancy D. Searby, Program Manager, Capacity Building, NASA Headquarters
- Lawrence Friedl, Senior Engagement Officer, Earth Science Division NASA Headquarters
- Dan Irwin, Global Program Manager, SERVIR, NASA Marshall Space Flight Center
- Ashutosh Limaye, Chief Scientist, SERVIR, NASA Marshall Space Flight Center
- Tony Kim, Project Manager, SERVIR, NASA Marshall Space Flight Center
- Robert Griffin, Associate Professor, Atmospheric and Earth Science, University of Alabama in Huntsville and University Program Lead, SERVIR Program
- Pete Epanchin, Senior Climate Adaptation and Resilience Advisor, USAID Bureau for Resilience, Environment and Food Security (REFS)
- Chelsea Kay, Climate Adaptation Advisor, USAID REFS
- Evan Notman, Senior Climate and Land Use Officer, USAID REFS
- Janet Nackoney, Land and Resource Governance Division, USAID REFS
- Aminata Diarra, Biodiversity and Climate Change Advisor, USAID West Africa
- Napak Tesprasith, Project Management Specialist (Sustainable Development), USAID Regional Development Mission for Asia

For insights from SERVIR hubs in Southeast Asia, Hindu Kush Himalaya, West Africa, and Amazonia:
- Ekapol Sirichaovanichkarn, Monitoring and Evaluation Lead, ADPC / SERVIR Southeast Asia (SEA)
- Peeranan Towashiraporn, Chief of Party, ADPC / SERVIR SEA
- Pierre C. Sibiry Traore, User Engagement Lead, ICRISAT / SERVIR West Africa (WA)
- Paul Bartel, Chief of Party, ICRISAT / SERVIR WA
- Bako Mamane, Co-Chief of Party, ICRISAT / SERVIR WA
- Foster Mensah, CERSGIS / SERVIR WA
- Kidia Gelaye, Science and Data Lead, ICRISAT / SERVIR WA
- Birendra Bajracharya, Chief of Party, ICIMOD / SERVIR Hindu Kush Himalaya (HKH)
- Rajesh Bahadur Thapa, Senior Remote Sensing and Geoinformation Specialist, ICIMOD / SERVIR HKH
- Carlos Gasco, Program Director, CIAT / SERVIR-Amazonia
- Brian Zutta, Science & Data Lead, SIG, SERVIR-Amazonia

For ideas and questions channeling projects related to the SERVIR Applied Sciences Team (AST):
- Naiara Pinto, Landscape Ecologist, NASA Jet Propulsion Laboratory / SERVIR AST PI
- Ben Zaitchik, Professor, John Hopkins University / SERVIR AST PI
- Robert Kennedy, Professor, Oregon State University / SERVIR AST PI

For broader valuation perspectives from previous and upcoming NASA-sponsored work:
- Shanna N. McClain, Program Manager, Disasters, NASA Headquarters
- Yusuke Kuwayama, Assistant Professor, University of Maryland Baltimore County / Ex-Director VALUABLES

For additional input and reviews from the SERVIR SCO:

Basu et al. 2024. Serving economic prosperity: economic impact assessments…

- Kian Schuerman, SERVIR Intercontinental Coordinator
- Ankit Joshi, SERVIR Asia Coordinator
- Phoebe Oduor, SERVIR Africa Coordinator
- Brent Roberts, SERVIR Theme Lead, Weather & Climate Resilience
- Betzy Hernández Sandoval, SERVIR Regional Science Coordination Lead - Central America
- Lauren Carey, SERVIR Regional Science Associate - Central America
- Katie Walker, SERVIR Science Associate - Ecosystem & Carbon Management
- Timothy Mayer, SERVIR Scientist - Data Science, Water Security, Weather & Climate Resilience
- Jacob Abramowitz, SERVIR Science Associate - Ecosystem & Carbon Management
- Vanesa Martin, SERVIR Regional Science Coordination Lead - Amazonia
- Stephanie Jiménez, SERVIR Regional Science Associate - Amazonia
- Meryl Kruskopf, SERVIR Science Associate - Water Security and Air Quality & Health

For insights from EO projects that were assessed in prior SERVIR performance evaluations:
- Africa Flores-Anderson, Research Physical Scientist, NASA Marshall Space Flight Center

For collaboration on the GEOGloWS economic impact assessment:
- Angelica Gutierrez, NOAA
- Amber Kremer, AmeriGEO

Funding for this work was provided through the cooperative agreement 80MSFC22N0004 between NASA and UAH, and through the cooperative agreement 80MSFC17M0022 between NASA and USRA. SERVIR is a joint NASA- and USAID-led program.

# APPENDIX

**Key Concepts**

EIA encompasses a variety of methods. The appropriate application of each method requires an understanding of key terms and concepts. Definitions of the key concepts below should help familiarize readers with the necessary background on analytical approaches discussed in the preceding sections.

**Bayesian approach** - This approach refers to application of Bayesian statistics and methods to analyze the economic effects of various interventions, policies, or events. Bayesian analysis starts with prior beliefs or prior information about the parameters of interest. In economic impact analysis, these parameters might include the effects of a policy on employment, wages, consumer spending, etc. The likelihood function represents the probability of observing the data given specific parameter values. Bayesian inference combines the prior information with the likelihood function to obtain the posterior distribution of the parameters. This posterior distribution represents updated beliefs about the parameters based on both prior knowledge and observed data. The posterior distribution allows for the estimation of the parameters of interest, along with measures of uncertainty (e.g., credible intervals) (Rens et al., 2021).

**Bayesian network (BN)** - A Bayesian network (or Bayes net) is a graphical model that represents probabilistic relationships among a set of variables. It is composed of nodes representing variables and directed edges representing dependencies between the variables. This Bayesian network can be used to calculate probabilities of certain events given evidence about other events (Chen & Pollino, 2012).

**Causal mediation analysis** - Causal mediation analysis is a statistical method used to investigate the mechanisms through which an independent variable influences a dependent variable. It aims to understand the underlying causal pathways by which the independent variable affects the outcome, often through one or more intermediate variables, known as mediators (Raymond & Dustin, 2012).

**Cost-benefit analysis (CBA)** - This is a method of economic analysis that places a value on inputs (costs) and outputs (benefits). It can account for personal valuations and how much that value changes depending upon uncertainties and risks (Quah & Toh, 2012).

**Counterfactual -** Counterfactual evaluation involves comparing the outcomes of individuals who have benefited from a policy or program (the "treated group") with those of a similar group who have not been exposed to the policy or program (the "comparison/control group"). This method helps determine what would have happened to the treated group if they hadn't been exposed to the intervention, known as the counterfactual case. This approach is crucial for collecting evidence to assess whether policy objectives have been achieved efficiently, informing future intervention designs and budgetary decisions (Ferraro, 2009).

**Decision Models** - A decision model in economics refers to frameworks used for analyzing choices made by economic decision makers. The decision models include Consumer Choice Models, Production and Cost Models, Investment Decision Models, Game Theory Models and Macro-economic Decision Models. Overall, decision models in economics provide a structured framework for understanding and predicting the behavior of economic agents in different contexts, helping economists and policymakers make informed decisions.

**Difference-in-difference (DID)** - This approach is a quasi-experimental design often used by researchers to study causal relationships in public health settings. Comparison groups, sensitivity analysis, and robustness checks help to validate the assumptions made in any DID model (Coady et al., 2018). For example, DID could be used to compare the mortality rates of a community that receives air quality warnings versus a community that doesn't.

**Dynamic integrated climate economy (DICE) model** - The DICE model is designed to analyze the economics of global warming, integrating economic and climate factors to assess the economic impacts of climate policies. The structure and background of the DICE model serves as a comprehensive tool for policymakers to evaluate the long-term consequences of climate change mitigation strategies. The model can incorporate dynamic elements to simulate interactions between the global economy, energy production, greenhouse gas emissions, climate dynamics, and policy interventions over multiple time periods. The model can analyze the costs and benefits of different policy options, such as carbon pricing mechanisms and emission reduction targets, to promote informed policy decisions aimed at mitigating climate issues while sustaining economic growth (William, 1992).

**Expected Value of Perfect Information (EVPI) -** It is a concept used in decision analysis, economics, and statistics to measure the value of having complete and perfect information before making a decision. It represents the maximum amount a decision-maker would pay to know the actual state of nature in advance. EVPI helps in evaluating the benefit of eliminating uncertainty in decision-making processes (Felli & Hazen, 1997).

**Fixed effects (FE)** - This is a popular statistical model in which it is assumed that one true effect size underlies all studies in the analysis and that all differences in observed effects are due to sampling error. FEs are often used in panel data or longitudinal studies where observations are made on the same entities. This term is also used in the context of regression models (Borenstein et al., 2010).

**General Equilibrium Models (GEMs) -** General equilibrium models within economics focuses on determining prices and quantities across multiple interconnected markets simultaneously. This stands in contrast to partial equilibrium analysis, which only considers a single sector. General equilibrium models are characterized by their comprehensive scope, addressing constraints at both individual and systemic levels throughout the economy. Widely employed across various branches of economics, particularly in macroeconomics and international trade theory, these models are instrumental in understanding the broader implications of economic changes. Given that alterations in one sector can reverberate throughout the entire economic system, general equilibrium models serve to elucidate these repercussions, providing insights into the full consequences of policy adjustments (Peeters et al., n.d.).

**Integrated Assessment Model (IAMs)** - Integrated assessment models strive to offer actionable insights for addressing global environmental change and fostering sustainable development. They achieve this by quantitatively depicting critical processes within human and Earth systems, as well as their intricate interconnections. IAMs adopt an integrative approach, drawing upon diverse scientific disciplines to comprehensively describe both human activities and Earth's dynamics. The term "assessment" underscores their focus on generating decision-relevant information, even amidst significant uncertainties. IAMs have proven effective in supporting climate policy by providing projections of future greenhouse gas emissions and identifying mitigation options. They have also been instrumental in various environmental evaluations (*Integrated Assessment Models (IAMs) and Energy-Environment-Economy (E3) Models*, n.d.).

**Input-output analysis (IOA)** - Economists prioritize promoting innovation and efficiency to maximize growth, profits, and consumption, often depicting the economy as a circular flow of income between producers and consumers. This model highlights the interaction between money flows and physical goods and services. However, economists typically overlook the environment's role, treating resources as free and waste as externalities. Input-output economics, pioneered by Wassily Leontief in the 1930s, addresses this gap by explicitly incorporating resource inputs and waste generation into its models, making it well-suited for environmental analysis. Unlike other economic approaches, input-output analysis does not solely focus on growth and efficiency but also accommodates other objectives. (Raa, 2005).

**Life Cycle Assessment (LCA)** - In economics, a life cycle assessment model is a methodical approach employed to analyze the environmental ramifications linked with every stage of a product's life cycle, spanning from the extraction of raw materials to production, usage, and eventual disposal or recycling. This model takes into account diverse environmental aspects including energy usage, depletion of resources, emissions of pollutants, and generation of waste. The LCA model serves to evaluate the environmental sustainability of products or procedures and to guide decision-making concerning the allocation of resources, methods of production, and the formulation of policies. By quantifying the environmental impacts throughout the entire lifespan of a product, enterprises, governmental bodies, and scholars can pinpoint areas for enhancement, optimize the utilization of resources, and mitigate adverse environmental consequences (Norris, 2001).

**Linear regressions** - Regression studies the dependence of one variable on another. The goal of linear regression is to summarize observed data. For example, how does agricultural production change based on the number of farmers that have access to drought-resistant crops.

**Multinomial logit probability** - These models are used to estimate the probability of choosing a specific category from a set of mutually exclusive categories. It is used to analyze and predict individual choices and determine or select a particular alternative (So & Kuhfed, 1995).

**Net present value (NPV)** - NPV is used to study profitability of a project or investment over time. It is the difference in present value of cash inflows and the present value of cash outflows over a period. NPV is calculated by discounting all future cash flows to their present value (adjusting for factors like inflation) and then subtracting that value from the initial cost (Gallo, 2014).

**Participatory action research (PAR)** - PAR prioritizes the significance of experiential knowledge in addressing challenges stemming from unjust and harmful social structures, and in conceptualizing and implementing alternative solutions. PAR entails the active involvement and leadership of individuals directly impacted by these issues, who engage in action to foster transformative social change by conducting organized research to produce innovative insights. (Costa & Andreaus, 2021).

**Random utility model (RUM)** - A type of mathematical model used in economics to analyze and predict choices made by individuals amongst a set of mutually exclusive discrete alternatives. Its applicability lies mainly in marketing and environmental economics (Baltas & Doyle, 2000).

**Randomized control trials (RCT)** - RCTs are increasingly popular in the social sciences. In an RCT, participants are randomly assigned to different groups, with one group receiving an intervention (the treatment group) and another group receiving either a placebo or standard treatment (the control group).

Data collected from RCTs are analyzed using statistical methods to determine whether the intervention has a significant effect on the outcome measures. Common statistical techniques include t-tests, chi-square tests, and regression analysis (Stolberg et al., 2004).

**Real Option Value (ROV) -** The discounted cash flow method may not fully account for the uncertainty surrounding the future financial performance of a business, ownership stake, or security under valuation. In such cases, when the valuation objective necessitates considering the potential influence of the owner/operator on future financial outcomes, real option valuation theory emerges as a potent analytical framework. ROV analysis finds frequent application among corporate acquirers and other investors who prioritize understanding the intrinsic value of an investment rather than its market value. Unlike conventional financial option valuation, which typically focuses on public or private company stock options, warrants, grants, or rights, ROV theory applies option pricing principles more expansively. It serves as a strategic investment tool, enabling stakeholders to assess investment opportunities with greater clarity (Schweihs, 1999).

**Revealed preference** - This terminology describes an individual's personal preference based on their observed behavior or choices in various economic situations. This theory suggests that individuals reveal their preferences through the choices they make and the way they behave in a marketplace rather than explicitly disclosing them in the form of statements (Richter, 1966).

**Social cost** - This is the cost incurred by society due to consumption or production of any goods or services. For example, air pollution could be a social cost of using a car. It includes externality costs which are not borne by any individual or entity directly involved in the economic activity (Coase, 1960).

**Social cost benefit analysis (SCBA) -** SBCA extends the scope of traditional CBA by considering broader social impacts and effects. In addition to assessing monetary benefits and costs, SCBA also takes into account non-monetary impacts such as social welfare, equity, environmental quality, and other social dimensions. By involving a more comprehensive analysis, SBCA is more suitable for evaluating projects or policies with significant social implications (Harberger, 1978).

**Social returns on investment analysis (SROI)** - SRO analysis is a methodology used to assess the social, environmental, and economic value generated by an organization, program, project, or intervention. It seeks to measure the positive outcomes or impacts created by an initiative relative to the resources invested. SROI analysis provides a holistic framework for assessing the impact and value of social initiatives, helping organizations and funders make informed decisions about resource allocation, investment, and strategic priorities (Pathak & Dattani, 2014).

**Structured Expert Judgment (SEJ) -** Structured expert judgment is a systematic and methodical method utilized to collect and evaluate expert viewpoints regarding uncertain or intricate matters. Within SEJ, specialists are tasked with offering their assessments or judgments within a well-defined framework, often comprising predetermined procedures and protocols. This methodology enables the incorporation of various expert viewpoints and streamlines the amalgamation of individual judgments into a unified appraisal (Van Steen, 1992).

**Value of information (VOI) -** The VOI refers to the benefits or advantages obtained from possessing relevant and timely data or knowledge. This concept is applicable across various domains of science and technology (Lawrence, 1999) . For example, the advantage of receiving a flood warning.

**Value of statistical life (VSL)** - VSL is a concept in economics used to assign monetary value to the prevention of one statistical death. This measure helps policymakers assess benefits of regulations or policies to reduce mortality risks (Shanmugam, 2013).

**Willingness-to-pay (WTP)** - The maximum amount an individual or entity is willing to pay for a product or service. This concept is used to assess economic value associated with a product, good, or service and how it changes based on quantity and utility (Breidert et al., 2006). For example, if there are more frequent droughts, the utility of agricultural insurance would be higher and farmers' WTP for coverage may be higher.